%% file: Template.tex
\documentclass[9pt]{article}

\usepackage{pgfplots, pgfplotstable,balance}
\usepackage{hyperref}
\usepackage{enumitem}
\pgfplotsset{compat=1.18}
\usepackage{spconf,amsmath,graphicx,makecell}
\usepackage{booktabs}
\usepackage{graphicx}
\usepackage{tipa}
\usepackage{xspace}
\usepackage{multirow}
\usepackage{url}
\usepackage{comment}
\usepackage{float}
\usepackage{tikz}
\usepackage{cite}
\usepackage{setspace}

\def\ci95{{CI$_{95}$}\xspace}
\newcommand{\unmarkedfootnote}[1]{%
  \begingroup
  \renewcommand\thefootnote{}
  \footnote{#1}%
  \addtocounter{footnote}{-1}
  \endgroup
}

\title{Assessing speech quality metrics for evaluation\\of neural audio codecs under clean speech conditions}
\vspace{-1cm}
\begin{NoHyper}
\name{\begin{tabular}{c}Wolfgang Mack$^{1,*}$, Nezih Topaloglu$^{2,*}$, Laura Lechler$^{1}$, Ivana Bali\'{c}$^{1}$, \\Alexandra Craciun$^{1}$, Mansur Yesilbursa$^{1}$, Kamil Wojcicki$^{1}$\end{tabular}}
\vspace{-5pt}
\address{%
$^1$Collaboration AI, Cisco Systems, Inc.\\
$^2$Faculty of Engineering and Natural Sciences, Yeditepe University, Istanbul, Turkey\\[2pt] 
\{\emph{womack}, \emph{kamilwoj}\}@cisco.com
} 
\end{NoHyper}
\vspace{-20pt}
\begin{document}

\maketitle
\begin{abstract}
\vspace{-5pt}
\iffalse
Objective speech-quality metrics are widely used to evaluate codecs. Especially for neural codecs, it is often unclear which metrics give a reliable quality estimate. To determine reliability, we evaluated 45 objective metrics by correlating their values with subjective listening scores on clean speech under 17 codec conditions. Neural-based objective metrics such as \texttt{scoreq} or \texttt{utmos} exhibited the highest Pearson correlation to subjective scores. Analysis on different subjective quality ranges revealed that non-intrusive objective metrics tend to saturate for high subjective scores.
\else
Objective speech-quality metrics are widely used to assess codec performance. However, for neural codecs, it is often unclear which metrics provide reliable quality estimates. To address this, we evaluated 45 objective metrics by correlating their scores with subjective listening scores for clean speech across 17 codec conditions. Neural-based metrics such as \texttt{score\_q} and \texttt{utmos} achieved the highest Pearson correlations with subjective scores. Further analysis across different subjective quality ranges revealed that non-intrusive metrics tend to saturate at high subjective quality levels.
\fi

\end{abstract}

\vspace{-6pt}
\vspace{-10pt}
\section{Introduction}
\label{sec:intro}
\vspace{-10pt}

Rigorous assessment of perceived speech quality is essential for modern telecommunication systems. It underpins quality-of-experience~(QoE) assurance in operation and enables evidence-based selection of signal processing and machine learning components during development. Speech quality can be assessed using subjective (human listening) and/or objective (algorithmic) methods. 
Subjective tests, such as absolute category rating (ACR) (ITU-T P.800), which yields the mean opinion score (MOS), along with MUSHRA (ITU-R BS.1534) and BS.1116, are considered the gold standard for evaluating perceptual quality.
These must be executed under standardized conditions and are generally time consuming and costly. Objective methods, in contrast, are fast, inexpensive, and amenable to continuous (online) monitoring. While classical metrics are often intrusive (reference-based) (e.g.~\texttt{pesq} 
\cite{rix_perceptual_2001}, \texttt{polqa}\cite{beerends_perceptual_2013}, and \texttt{warpq} \cite{jassim2021warpq}), there has been a growing interest  \cite{voicemos2022, voicemos2023, voicemos2024} in developing non-intrusive methods that estimate quality directly from the test signal alone, such as \texttt{dnsmos} \cite{Reddy22_dns_overall}, \texttt{nisqa} \cite{mittag_nisqa_2021-1} and \texttt{utmos} \cite{saeki2022utmos}. \begin{NoHyper}
\unmarkedfootnote{* Equal contribution.}
\end{NoHyper}

Neural audio codecs have recently emerged as a transformative alternative to traditional DSP-based codecs, achieving intelligible and often high-quality speech transmission even at ultralow bitrates ($<$$2$kbps) \cite{Wu2023audiodec,defossez_high_2022,kumar2023DAC,gu2024esc,siuzdak2024snac,Liu2024semanticodec,li2025dualcodec,parker2024stablecodec}. Their adoption is accelerating in bandwidth-constrained applications such as large-scale conferencing, cloud gaming, XR/VR, and on-device communication \cite{xin2024bigcodec,defossez2024moshi}. However, such codecs can exhibit generative artifacts including temporal inconsistencies, perceptual hallucinations, and unnatural timbre shifts, which differ fundamentally from the distortions seen in traditional codecs \cite{lechler2025}. As a result, widely used objective metrics, which were originally optimized for perceptual fidelity and assessment of DSP-introduced artifacts, may correlate poorly with human preference of speech quality for neural codecs \cite{lechler2025,shi2025versa}. This mismatch undermines both codec development (metric-driven model selection) and monitoring  of operational quality.


Despite the rapid development of neural codecs, there is no comprehensive large-scale study that systematically compares diverse objective metrics, intrusive and non-intrusive alike, against crowdsourced listening tests across both low-bitrate and ultralow-bitrate operating points.
Prior comparisons \cite{10694655,lechler2025} are restricted in both the number of objective metrics and codec conditions.
In contrast, the present work mitigates the above knowledge gap by benchmarking a large suite of objective metrics against listener judgments.
Specifically, we report on the evaluation of 45 objective metrics under 17 codec conditions and corresponding crowdsourced listening test results. 
The listening tests leverage the MUSHRA-1S \cite{lechler2025mushra1s} methodology capable of resolving subtle quality differences at the higher end of the quality scale more effectively than achievable by ACR tests \cite{Koester2015ACRContinuous}.
More importantly, the listening tests are conducted at 24~kHz, which aligns with the sampling rate of many neural codecs.
We then analyze rank and linear correlations to derive practical guidelines for selecting metrics that are both discriminative for fast codec comparison and robust for monitoring service quality.
%
%

Our main contributions include a unified, large-scale comparison of intrusive and non-intrusive metrics on modern neural codecs, based on MUSHRA-1S listening tests \cite{lechler2025mushra1s}. Additionally, this work introduces a reproducible evaluation protocol designed to enable efficient and repeatable estimation of metric-perception alignment at scale. Finally,  empirically grounded recommendations for selecting objective metrics are offered. These are aimed at helping practitioners choose metrics that better reflect human perception, thereby speeding up model development and maintaining confidence in quality outcomes for neural codecs.

\section{Methods}\vspace{-10pt}
This section introduces the signal model, details the computation of correlations between objective and subjective scores, and outlines the evaluation methods employed.
\vspace{-10pt}
\subsection{Fundamentals}
\vspace{-5pt}
We consider a set $\mathcal{X}$ of $N$ discrete time-domain speech signals, 
i.e.,
\begin{equation}
\mathcal{X} = \{x_1, \ldots,x_N\}.
\end{equation}
A set of $C$ neural audio codec conditions (codecs and bitrates)
$\mathcal{A} = \{\mathcal{A}_1, \ldots, \mathcal{A}_C \}$ are applied to speech in $\mathcal{X}$, yielding corresponding $C$ processed sets $\mathcal{P} =\{\mathcal{X}_1, \ldots, \mathcal{X_C}\}$, where $\mathcal{X}_C = \{x_1^C, \ldots, x_N^C\}$ contains the speech processed by $\mathcal{A}_C$.

All files in $\mathcal{P}$ are evaluated using objective and subjective metrics, leading to a total of $C\cdot N$ scores per metric, $N$ for each codec condition. For each objective metric, we compute the Pearson \cite{pearson1896vii}, Spearman \cite{spearman1961proof} and Kendall \cite{kendall1948rank} correlations of the $C\cdot N$ scores against their subjective counterparts. 
Reporting on all three correlations accounts for different assumptions~(linearity vs.~monotonicity; raw values vs.~ranks) and guards against outliers and ties: Pearson quantifies linear association, Spearman applies Pearson to ranks to capture general monotonic trends and handle ordinal data, while Kendall compares concordant/discordant pairs, offering a probabilistic interpretation and stronger tie handling.

\vspace{-10pt}
\subsection{Subjective Evaluation and Objective Metrics}
\vspace{-5pt}
For subjective evaluation, we adopt the recently proposed MUSHRA-1S variant of the MUSHRA test, which includes a single test condition \cite{lechler2025mushra1s}. MUSHRA-1S was shown to be more scalable with respect to the number of conditions than regular MUSHRA tests and to have a more fine-grained evaluation scale than ACR tests \cite{Koester2015ACRContinuous}. Scalability in terms of condition-number is crucial in our tests, as we compare many objective scores with many codec conditions. In MUSHRA-1S, participants are presented with a single condition plus anchor and reference. The results of multiple tests can be merged to achieve a reliable comparison of a large number of conditions, something not achievable with a single MUSHRA test and challenging to achieve with multiple MUSHRA tests. 

The objective metrics we consider range from traditional signal-processing metrics such as \texttt{pesq} to state-of-the-art neural-based metrics such as \texttt{utmos} or \texttt{scoreq}. We compare both intrusive (with reference) and non-intrusive metrics. 

\vspace{-10pt}
\section{Experiments} 
\vspace{-5pt}
\begin{table}[t]
\centering
\caption{Neural audio codecs evaluated in this study, sorted alphabetically. }
\begin{tabular}{lccccc}
\hline
\textbf{Codec name} & \makecell{\textbf{Bitrates} \\ \textbf{(kbps)}} & \makecell{\textbf{Model $F_s$} \\ \textbf{(kHz)}} & \textbf{Year}  \\
\hline
Audiodec \cite{Wu2023audiodec} & 6.4 & 24  & 2023 \\
BigCodec \cite{xin2024bigcodec} & 1.0 & 16  & 2024 & \\

DAC \cite{kumar2023DAC} & 1.3, 5.3, 8.0 & 24  & 2023 & \\
DAC Tiny$^{*}$ \cite{gu2024esc} & 8.0 & 16 & 2024 & \\
DualCodec$^{**}$ \cite{li2025dualcodec}& 0.9 & 24 & 2025 & \\
Encodec \cite{defossez_high_2022} & 1.5, 6.0 & 24  & 2022 \\
ESC \cite{gu2024esc} & 1.5, 6.0 & 16  & 2024 \\

Internal Codec$^{***}$ & 1.0, 6.0 & 16 & 2024 & \\

Mimi \cite{defossez2024moshi}& 1.1 & 24 & 2024 & \\

Semanticodec \cite{Liu2024semanticodec}& 1.4 & 16  & 2024 & \\
SNAC \cite{siuzdak2024snac} & 1.0 & 24 & 2024 \\

StableCodec \cite{parker2024stablecodec}& 1.0 & 16 & 2024 & \\

\hline
\end{tabular}
\\[2mm]
\small $^{*}$A lightweight 16 kHz variant from the DAC family, published by the ESC team \cite{gu2024esc}.\small $^{**}$Frame rate: 12.5 Hz, version: v1.\small $^{***}$An internal audio codec included to improve the coverage of the MUSHRA scale.
\label{tab:list_of_codecs}
\end{table}
We evaluated a total of 17 codec conditions with 45 objective metrics and MUSHRA-1S on 100 clean speech files from the low-resource audio coding (LRAC) challenge\footnote{https://lrac.short.gy} blind test set of Track 1. As a result of the subjective testing, the mean number of votes per file is approximately 7, following the suggestion in \cite{lechler2025mushra1s}.

Table~\ref{tab:list_of_codecs} provides an overview of the neural codecs and their respective bitrates. The codecs and bitrates were selected such that the audio quality covers large parts of the MUSHRA range (0--100). The selected codecs intentionally differ, e.g., in terms of network design and training, compute, latency, bitrate and spectral bandwidth, to help support a diverse range of output distortions. The codec sampling frequency depends on the system and is listed in Table~\ref{tab:list_of_codecs}. 

For the crowdsourced subjective tests, MUSHRA-1S was used. The anchor and reference were sampled at 24~kHz. The anchor was processed by Opus 6~kbps \cite{opus}. 

For the objective evaluation, we used the VERSA toolkit, a framework for a wide range of speech and audio quality metrics \cite{shi2025versa}. For intrusive metrics, the sampling frequency of the reference signals was adjusted to that of the codecs. In this investigation, we consider codecs with MUSHRA-1S scores in the range of [0, 50] as low quality, [50, 75] as medium quality, and [75, 100] as high quality. 
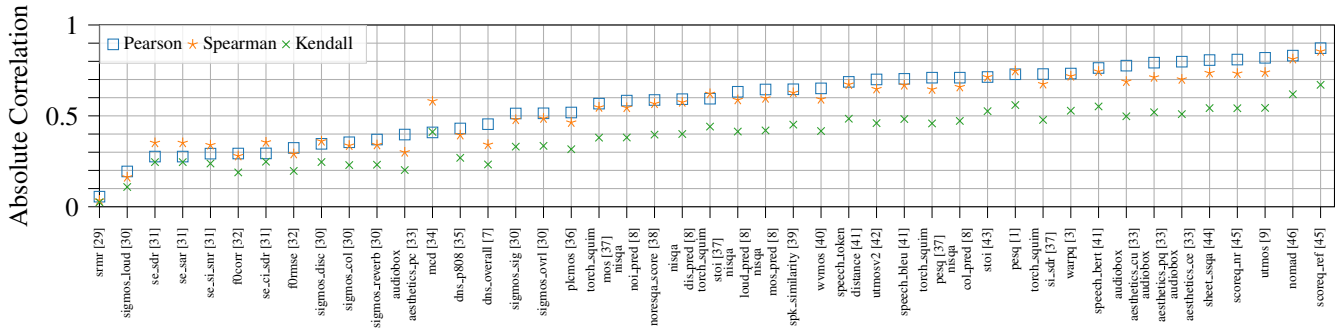
\begin{figure*}
\centering
\begin{NoHyper}
\input{plots/pearsonsummary} 
\end{NoHyper}
\vspace{-25pt}
\caption{Absolute correlation coefficients of objective metrics (from VERSA \cite{shi2025versa}) with subjective scores. The metrics are ordered from weakest to strongest  Pearson correlation. We evaluated 17 conditions with 100 samples each, resulting in a total of 1700 data points for the correlation analysis.}
\vspace{-10pt}
\label{fig:pearsonsummary}
\end{figure*}

\vspace{-10pt}
\section{Correlation Analysis}\vspace{-5pt}
\label{sec:typestyle}

In this section, we assess the correlation of objective scores with subjective results for different codec conditions.

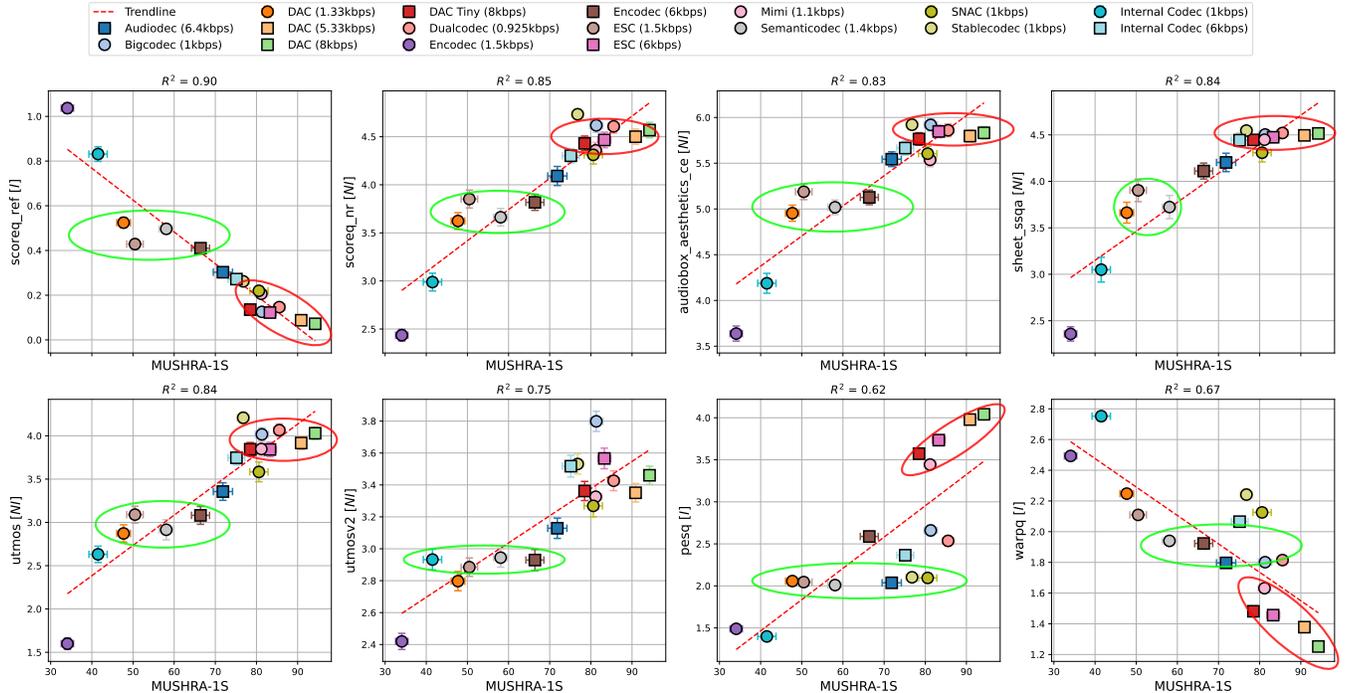
\begin{figure*}[t]
  \centering
    \begin{tikzpicture}
        
        \node[anchor=south west, inner sep=0] (img) at (0,0) 
            {\includegraphics[width=\textwidth]{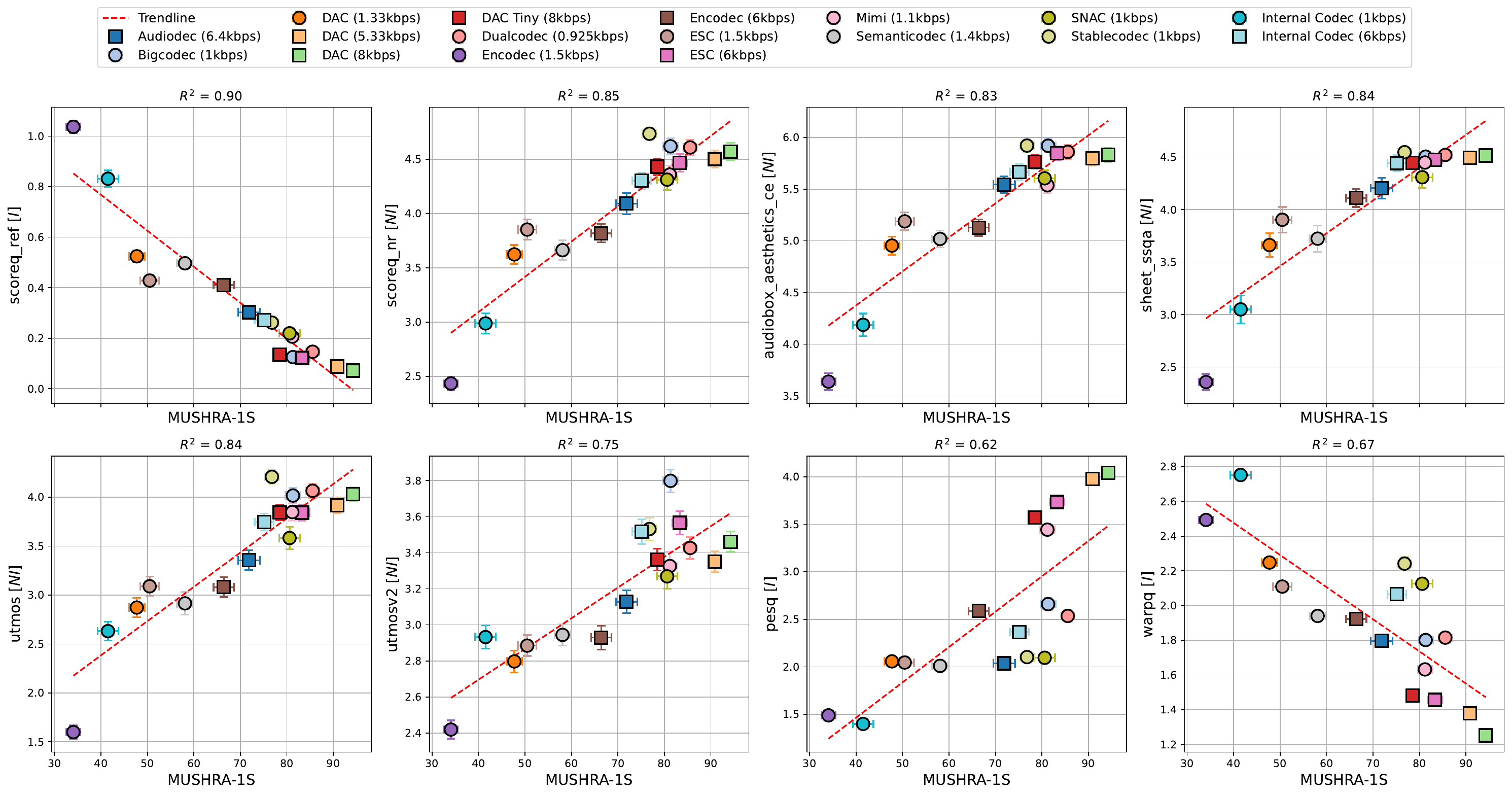}};
        \begin{scope}[x={(img.south east)},y={(img.north west)}]
    
            \draw[green, opacity=0.8, thick]
                (0.12,0.25) ellipse (0.05 and 0.033);
    
            \draw[red, opacity=0.8, thick]
                (0.21,0.37) ellipse (0.04 and 0.03);
                
            \draw[green, opacity=0.8, thick]
                (0.11,0.66) ellipse (0.06 and 0.035);
    
            \draw[red, opacity=0.8, thick, 
            rotate around={-30:(0.21,0.55)}] 
                (0.21,0.55) ellipse (0.04 and 0.03);
    
            \draw[green, opacity=0.8, thick]
                (0.36,0.2) ellipse (0.06 and 0.02);
    
    
            \draw[green, opacity=0.8, thick]
                (0.37,0.693) ellipse (0.05 and 0.03);
    
            \draw[red, opacity=0.8, thick]
                (0.45,0.8) ellipse (0.04 and 0.025);
    
            \draw[green, opacity=0.8, thick]
                (0.64,0.17) ellipse (0.08 and 0.025);
    
            \draw[red, opacity=0.8, thick,rotate around={33:(0.71,0.37)}] 
                (0.71,0.37) ellipse (0.045 and 0.023);
    
            \draw[green, opacity=0.8, thick]
                (0.62,0.70) ellipse (0.06 and 0.035);
    
            \draw[red, opacity=0.8, thick]
                (0.71,0.81) ellipse (0.045 and 0.023);
    
            \draw[green, opacity=0.8, thick]
                (0.91,0.22) ellipse (0.06 and 0.03);
    
            \draw[red, opacity=0.8, thick, rotate around={-42:(0.96,0.11)}] 
                (0.96,0.11) ellipse (0.048 and 0.03);
    
            \draw[green, opacity=0.8, thick]
                (0.855,0.70) ellipse (0.025 and 0.04);
    
            \draw[red, opacity=0.8, thick]
                (0.95,0.805) ellipse (0.045 and 0.024);
        \end{scope}
    \end{tikzpicture}
    \vspace{-18pt}
    \caption{Condition-wise plot of average objective vs.~subjective scores (merged MUSHRA-1S) with 95~\% confidence intervals (based on 100 files per condition) and linear regression line. The red ellipses mark codecs that obtain high scores both in MUSHRA-1S and the corresponding objective metric. The green ellipses indicate conditions that are scored very similarly by objective metric while having large spread in subjective score. Circle/circular markers depict ultra-low bitrates  (around 1 kbps), square markers depict low bitrates (around 6 kbps). Abbreviations I and NI specify intrusive and non-intrusive metrics, respectively. }
    \label{fig:detailedeval}
    \vspace{-10pt}
\end{figure*}

The absolute Pearson, Spearman, and Kendall correlations for each objective metric with MUSHRA scores are shown in Figure~\ref{fig:pearsonsummary}. The weakest correlations arise for reverberation-focused predictors (\texttt{srmr}, \texttt{sigmos\_reverb}) (which are largely irrelevant for clean speech), $f_{0}$-error metrics (\texttt{f0corr}, \texttt{f0rmse}), and waveform-matching measures (\texttt{se\_sdr}, \texttt{se\_sar}, \texttt{se\_si\_snr}), which is consistent with the generative, non–waveform-preserving nature of neural codecs. Midrange performance is observed for metrics such as \texttt{torch\_squim}, \texttt{sigmos\_sig}, \texttt{nisqa}, and \texttt{wvmos}. The highest Pearson correlations are obtained for the \texttt{scoreq} family (\texttt{scoreq\_ref}: $0.87$, \texttt{nomad}: $0.83$, \texttt{scoreq\_nr}: $0.81$), \texttt{audiobox\_aesthetics\_ce} ($0.80$), \texttt{sheet\_ssqa} ($0.81)$, and \texttt{utmos} ($0.82$).  For the intrusive metrics with the strongest correlations (\texttt{scoreq\_ref} and \texttt{nomad}), Pearson, Spearman, and Kendall produced mostly identical rankings of methods, indicating that the associations are mostly monotonic and approximately linear, with little influence from outliers or ties. The best-correlated non-intrusive metrics (\texttt{utmos}, \texttt{scoreq\_nr}) show much smaller Spearman and Kendall correlation coefficients compared to their Pearson correlation. This phenomenon suggests ranking deficiencies of these metrics and will be analyzed in more detail in Figure \ref{fig:detailedeval}. \texttt{scoreq\_ref} achieves the highest correlation in Pearson, Spearman, and Kendall.  It is worth to note that all the metrics that exhibit Pearson correlation higher than 0.8 are based on neural networks. \texttt{warpq} and 
\texttt{pesq} perform best among the classical baselines, with a Pearson correlation of $0.73$.
\begin{figure}[t]
  \centering
  \includegraphics[width=0.975\linewidth]{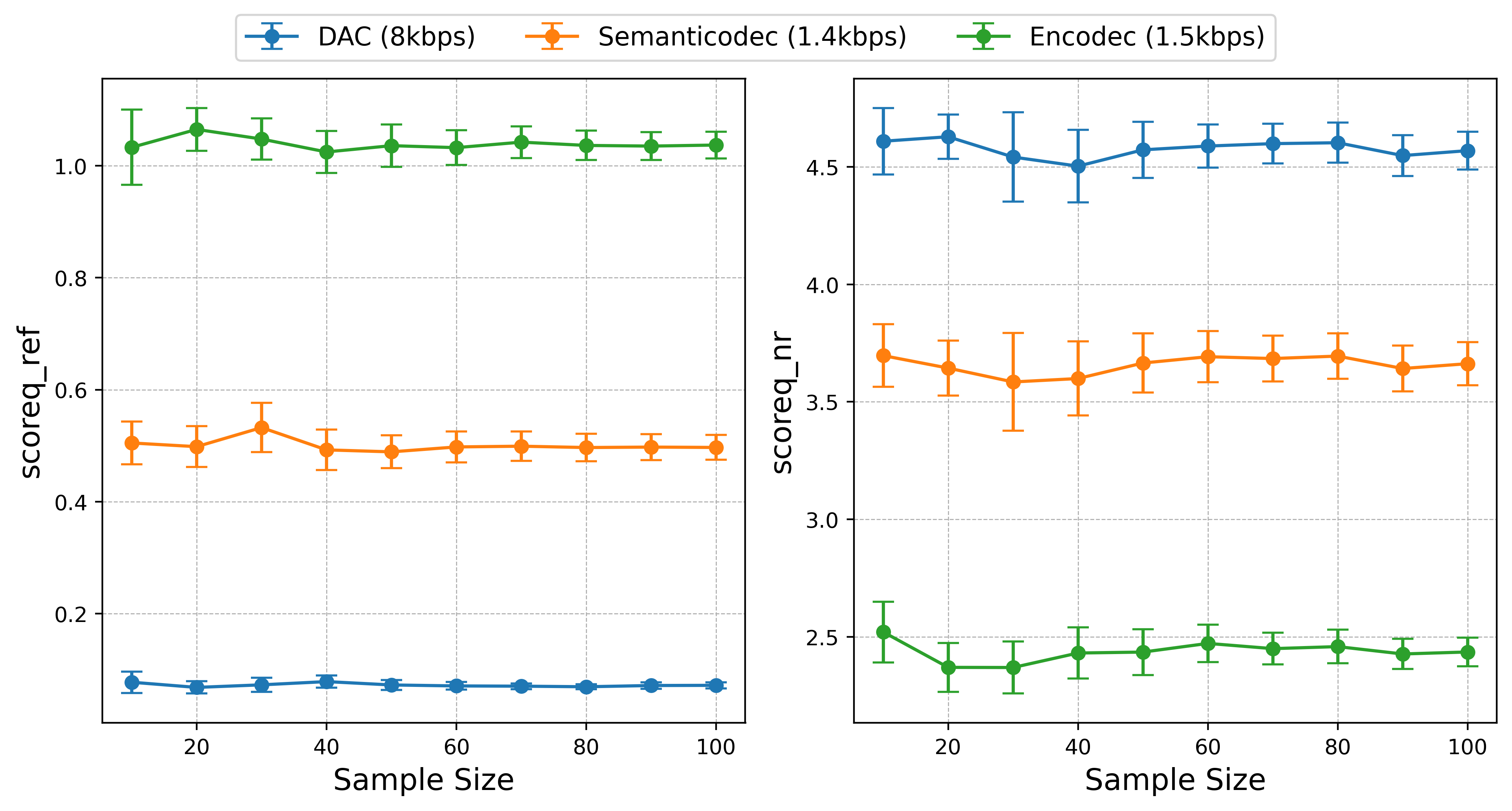}
    \vspace{-0.5cm}
  \caption{Mean and confidence intervals of two selected objective metrics over a subset of audio files for a high-, medium-, and low-quality codec condition.}
  \label{fig:confinterval}
  \vspace{-0.5cm}
\end{figure}

To investigate the effect of the number of audio files on the objective scores, an evaluation of the confidence intervals over a random subset of audio files is depicted in Figure \ref{fig:confinterval} for selected codecs and metrics. As representatives of intrusive and non-intrusive metrics, we report the mean values and confidence intervals of \texttt{scoreq\_ref} and \texttt{scoreq\_nr} for a high-, medium-, and low-quality codec condition: DAC-8kbps, Semanticodec-1.4kbps, Encodec-1.5kbps, respectively. As expected, increasing the sample size decreases the confidence intervals. Overall, the confidence intervals of the intrusive method are smaller than those of the non-intrusive method. The reference seems to stabilize the scores. This behavior is especially visible for the high-quality codec, where the confidence interval of the intrusive method is very small even for small sample sizes. For the non-intrusive method, the confidence intervals of all codec evaluations are in a comparable range for various sample sizes.
Similar results were found for other selected intrusive/non-intrusive methods. A jump in the mean values from sample size 10 to 20/30 is observed for low and medium quality codecs. This suggests using more samples for such codecs. The mean stabilizes for higher sample sizes. For ranking codecs, the sample size needs to be determined based on significance-testing between the scores.

To analyze the codec rankings in more detail, Figure~\ref{fig:detailedeval} shows objective vs.~subjective scores for high-correlation metrics across all codecs. Since \texttt{scoreq} and \texttt{nomad} are very similar, we do not report \texttt{nomad}, but two commonly used metrics: \texttt{pesq} and \texttt{utmosv2}. The plots in Figure~\ref{fig:detailedeval} show the mean and confidence intervals together with a regression line of objective vs.~subjective scores for individual codecs and metrics. Within the \texttt{utmos} family, \texttt{utmos} consistently outperforms \texttt{utmosv2}. Note that we observed run-to-run variability for \texttt{utmosv2}, possibly due to internal random frame selection (not currently exposed in VERSA). Overall,    \texttt{scoreq\_ref} performs best and follows an almost linear trend across the majority of codecs (high reliability). \texttt{scoreq\_nr} performs slightly worse, especially at high MUSHRA scores, where \texttt{scoreq\_nr} gives almost identical values for a wide range of subjective scores. 

To further investigate objective metric saturation for high MUSHRA-1S scores, we marked that range for selected metrics in Figure~\ref{fig:detailedeval} with red ellipses. The intrusive metrics \texttt{scoreq\_ref}, \texttt{pesq}, \texttt{warpq} show a near-linear correlation, while non-intrusive metrics such as \texttt{scoreq\_nr}, \texttt{ce} from audiobox, \texttt{sheet\_ssqa}, and \texttt{utmos} exhibit near constant objective scores here. We hypothesize that the saturation is caused by MOS-based training objectives of the non-intrusive methods. The target MOS is obtained via ACR tests which have a coarse 5-point scale per answer. Though multiple answers can lead to a continuous range, high-quality conditions might consistently be rated as 5, leading to difficulties in discriminating between them \cite{lechler2025mushra1s}. Consequently, non-intrusive metrics trained with MOS scores are less sensitive for comparing high-quality codecs. This finding is consistent with the drop of the Spearman and Kendall correlations with respect to the Pearson correlation reported earlier for top non-intrusive metrics. The drop is likely caused by score clustering in the high-MUSHRA range, leading to ranking issues from tied scores. Alternatively, it is arguable whether codecs with such high MUSHRA scores have quality differences, or if the intrusive objective metrics and the MUSHRA-1S tests evaluate non-quality-degrading differences to the reference signal. 

Another interesting observation is that for a wide range of MUSHRA scores, some of the corresponding objective scores are nearly constant in the low and medium ranges. This is marked by green ellipses in Figure~\ref{fig:detailedeval}.  All objective metrics exhibit this behavior to some extent, which is most prominent for \texttt{pesq}, \texttt{utmosv2}, and \texttt{warpq}. This behavior suggests that human perception is sensitive to certain codec distortions, which the objective metrics fail to capture. For \texttt{pesq}, this is a known phenomenon \cite{pesqaterian, torcoli_objective_2021}. For the evaluated codecs, the smallest MUSHRA range in the green ellipses that is mapped to nearly identical values is exhibited by \texttt{sheet\_ssqa}, where codecs spread about 10 MUSHRA points. The widest spread of around 35 MUSHRA points is observed for \texttt{pesq}. 

Overall, we recommend using \texttt{scoreq}, \texttt{utmos} and \texttt{ce} from audiobox, along with \texttt{sheet\_ssqa} for codec evaluation, for cases expected to fall within the low to medium MUSHRA score range. For codecs that are likely to operate in the high MUSHRA range, we recommend the use of intrusive metrics such as \texttt{scoreq\_ref}.

\vspace{-10pt}

\section{Conclusion}

\vspace{-5pt}
\label{sec:conclusion}
We ran listening tests across 17 codec conditions on clean speech and compared the resulting scores with 45 objective metrics. The strongest correlations were obtained by neural approaches, in particular \texttt{scoreq} variants, the audiobox metric 
\texttt{ce}, \texttt{sheet\_ssqa}, and \texttt{utmos}. The strongest correlation was achieved by \texttt{scoreq\_ref} with a Pearson correlation of $-0.87$. Among classical objective metrics, \texttt{warpq} and \texttt{pesq} performed best, with an absolute Pearson correlation of $0.73$. Several non-intrusive, MOS-trained metrics saturated at high subjective quality, thus weakening rankings. Intrusive metrics remained discriminative, with smaller confidence intervals. 
As a practical guideline for codec evaluation, we proposed to use \texttt{scoreq}/\texttt{utmos}/\texttt{ce}/\texttt{sheet\_ssqa} for low--medium quality conditions, and intrusive metrics such as \texttt{scoreq\_ref} for higher quality conditions. The confidence interval analysis reveals a shrinkage with sample size (the confidence intervals of intrusive metrics are smaller). The selection of the sample size should be based on the desired confidence interval width. Future work includes analysis of the influence of noise and reverberation on objective metrics.

\vfill\pagebreak

\newpage

\scriptsize
\let\oldthebibliography\thebibliography
\let\endoldthebibliography\endthebibliography
\renewenvironment{thebibliography}[1]{%
  \begin{oldthebibliography}{#1}%
    \setlength{\itemsep}{5pt}%
    \setlength{\parskip}{0pt}%
    \setlength{\parsep}{0pt}%
}{\end{oldthebibliography}}
\makeatother
\bibliographystyle{IEEEbib}
\balance

\bibliography{refs}
\end{document}

%% file: plots/pearsonsummary.tex
\begin{tikzpicture}

\definecolor{darkgray176}{RGB}{176,176,176}
\definecolor{darkorange25512714}{RGB}{255,127,14}
\definecolor{forestgreen4416044}{RGB}{44,160,44}
\definecolor{lightgray204}{RGB}{204,204,204}
\definecolor{steelblue31119180}{RGB}{31,119,180}

\begin{axis}[
legend cell align={center},
legend style={  font=\scriptsize,
legend columns=-1,
  fill opacity=0.8,
  draw opacity=1,
  text opacity=1,
  at={(0.,1.)},
  anchor=north west,
  draw=lightgray204
},
  width=18cm,
  height=4cm,
tick align=outside,
tick pos=left,
title={},
x grid style={darkgray176},
ylabel={Absolute Correlation},
xmajorgrids,
xmin=0, xmax=44.5,
xtick style={color=black},
ytick={0,0.1, 0.2, 0.3, 0.4, 0.5, 0.6, 0.7, 0.8, 0.9, 1},
yticklabels = {0,,,,,0.5, ,,,,1},
xtick={0,1,2,3,4,5,6,7,8,9,10,11,12,13,14,15,16,17,18,19,20,21,22,23,24,25,26,27,28,29,30,31,32,33,34,35,36,37,38,39,40,41,42,43,44},
xticklabel style={font=\tiny,
rotate=90.0},
align=center,
xticklabels={srmr \cite{Falk2010_srmr},
sigmos\_loud \cite{naderi2023sigmos},
se\_sdr \cite{lu22c_interspeech},
se\_sar \cite{lu22c_interspeech},
se\_si\_snr \cite{lu22c_interspeech},
f0corr \cite{Hayashi2020_f0},
se\_ci\_sdr \cite{lu22c_interspeech},
f0rmse \cite{Hayashi2020_f0},
sigmos\_disc \cite{naderi2023sigmos},
sigmos\_col \cite{naderi2023sigmos},
sigmos\_reverb \cite{naderi2023sigmos},
audiobox \\aesthetics\_pc \cite{tjandra2025metaaudioboxaesthetics},
mcd \cite{Kubichek93_mcd},
dns\_p808 \cite{Naderi2020AnOS},
dns\_overall \cite{Reddy22_dns_overall},
sigmos\_sig \cite{naderi2023sigmos},
sigmos\_ovrl \cite{naderi2023sigmos},
plcmos \cite{diener2023plcmosdatadrivennonintrusive},
torch\_squim\\mos \cite{Kumar23_torchsquim},
nisqa\\noi\_pred \cite{mittag_nisqa_2021-1},
noresqa\_score \cite{manocha2021noresqa},
nisqa\\dis\_pred \cite{mittag_nisqa_2021-1},
torch\_squim\\stoi \cite{Kumar23_torchsquim},
nisqa\\loud\_pred \cite{mittag_nisqa_2021-1},
nisqa\\mos\_pred \cite{mittag_nisqa_2021-1},
spk\_similarity \cite{jung2024_spk_similarity},
wvmos \cite{Andreev_2023_wvmos},
speech\_token \\distance \cite{Saeki2024_speech_bert_score},
utmosv2 \cite{baba2024utmosv2},
speech\_bleu \cite{Saeki2024_speech_bert_score},
torch\_squim \\pesq \cite{Kumar23_torchsquim},
nisqa \\col\_pred \cite{mittag_nisqa_2021-1},
stoi \cite{Taqal2010_stoi},
pesq \cite{rix_perceptual_2001},
torch\_squim \\si\_sdr \cite{Kumar23_torchsquim},
warpq \cite{jassim2021warpq},
speech\_bert \cite{Saeki2024_speech_bert_score},
audiobox\\aesthetics\_cu \cite{tjandra2025metaaudioboxaesthetics},
audiobox\\aesthetics\_pq \cite{tjandra2025metaaudioboxaesthetics},
audiobox\\aesthetics\_ce \cite{tjandra2025metaaudioboxaesthetics},
sheet\_ssqa \cite{huang2024sheetssqa},
scoreq\_nr \cite{ragano2025scoreq},
utmos \cite{saeki2022utmos},
nomad \cite{ragano_nomad_2024},
scoreq\_ref \cite{ragano2025scoreq}
},
y grid style={darkgray176},
ymajorgrids,
ymin=0, ymax=1,
ytick style={color=black}
]
\addplot [draw=steelblue31119180, fill=steelblue31119180, mark=square, only marks]
table{%
x  y
0 0.0554276922156213
1 0.194172858057867
2 0.276045294462808
3 0.276045294462808
4 0.292216982025234
5 0.292793049513271
6 0.293424193350956
7 0.323680945780378
8 0.346210803629451
9 0.354776209667093
10 0.369813205483166
11 0.397171138897035
12 0.408942254779085
13 0.430634042393542
14 0.454351041120847
15 0.513081631398784
16 0.514079329661228
17 0.518509273310312
18 0.567204809862377
19 0.583913823476791
20 0.587277781729617
21 0.591817006642543
22 0.594435461601427
23 0.632517263912511
24 0.645144812511786
25 0.646292397269565
26 0.651254014834158
27 0.686872594442568
28 0.700423698594791
29 0.703580512831041
30 0.710044143028767
31 0.710074197411625
32 0.713733142342697
33 0.729042989731118
34 0.729988758538546
35 0.732760184453384
36 0.763284315947816
37 0.776244312467816
38 0.792691146638884
39 0.798091692203417
40 0.807043246738556
41 0.809723593655427
42 0.8193621032733
43 0.831416903298861
44 0.872512776306464
};
\addlegendentry{Pearson}
\addplot [draw=darkorange25512714, fill=darkorange25512714, mark=star, only marks]
table{%
x  y
0 0.0352413773251742
1 0.162391150711313
2 0.351792433291341
3 0.351792433291341
4 0.339421115649818
5 0.279345268602479
6 0.354648083111417
7 0.290509502039768
8 0.360737749580972
9 0.336711321163551
10 0.340049723164683
11 0.299804915172873
12 0.580907544965765
13 0.394440334670395
14 0.341358685136693
15 0.4787657970552
16 0.485925588835788
17 0.463688191108618
18 0.546935404072656
19 0.544935873248256
20 0.567062644636343
21 0.574827123045403
22 0.621657491814957
23 0.587882945147744
24 0.596655292222075
25 0.627200575020833
26 0.591355985542313
27 0.672463720986008
28 0.647642417032497
29 0.66866637669519
30 0.64551896564337
31 0.658882500450848
32 0.713465211153074
33 0.747241991361099
34 0.674751448996685
35 0.718263730586814
36 0.744375168411859
37 0.688600002981226
38 0.711888833933733
39 0.700736158137199
40 0.736011291958593
41 0.732825367146386
42 0.738635627197742
43 0.812659349147526
44 0.855011838095945
};
\addlegendentry{Spearman}
\addplot [draw=forestgreen4416044, fill=forestgreen4416044, mark=x, only marks]
table{%
x  y
0 0.0243230491210042
1 0.108656359883311
2 0.245628747190292
3 0.245628747190292
4 0.237834165234285
5 0.188896182120778
6 0.247953467364912
7 0.197122305184186
8 0.24605458931041
9 0.229379187989918
10 0.231304713791118
11 0.201852078766467
12 0.411693606532844
13 0.269081808698785
14 0.232698804358355
15 0.330775794734388
16 0.334970425395027
17 0.316621079072547
18 0.380137445708864
19 0.380928685341834
20 0.396699717559697
21 0.40024055929297
22 0.440971351995767
23 0.413780544529127
24 0.419905646466052
25 0.451771996010377
26 0.416500005660488
27 0.484403963711427
28 0.460103172797465
29 0.482599850989348
30 0.45843024562026
31 0.472233494732679
32 0.525900274207637
33 0.559550011684712
34 0.477864708536741
35 0.528329327739202
36 0.551816116003374
37 0.497696665678293
38 0.519975389296883
39 0.509576561253686
40 0.542834855002308
41 0.542043896727037
42 0.543847954364911
43 0.619041096387269
44 0.670768283090376
};
\addlegendentry{Kendall}

\end{axis}

\end{tikzpicture}